\documentclass[a4paper,11pt]{article}
\usepackage{pos}
\usepackage{subcaption}
\usepackage{booktabs}  % 
\usepackage{gensymb}   % for symbles like \degree 

\title{Methods for statistical detection of GRBs in the context of the LST-CTAO}
\author*[a]{Léo Le Moigne}
\author[b]{Mathieu de Bony de Lavergne}
\author[a]{Armand Fiasson}
\author[a]{Edna Ruiz-Velasco}
\author[a]{David Sanchez}
\author[c]{Kenta Terauchi}

\affiliation[a]{Laboratoire d’Annecy de Physique des Particules (LAPP), Université Savoie Mont Blanc, CNRS/IN2P3, 9 Chemin de Bellevue, 74940 Annecy, France.}
\affiliation[b]{Centre de Physique des Particules de Marseille (CPPM), Aix–Marseille Université, CNRS/IN2P3, 163 Avenue de Luminy – Case 902, 13288 Marseille cedex 9, France.}

\affiliation[c]{Division of Physics and Astronomy,, Graduate School of Science, Kyoto University, Kitashirakawa‐Oiwakecho, Sakyo‐ku, Kyoto 606-8502, Japan.}

\emailAdd{lemoigne@lapp.in2p3.fr}

\abstract{Gamma-Ray Bursts (GRBs) afterglows are rapidly decaying signals that pose significant detection challenges, requiring improved methods to track their temporal evolution. In this study, we systematically compare various techniques for detecting GRB-like transient emissions at very high energies (VHE, >100 GeV). Our analysis includes time-dependent extension of the standard method (i.e., Li \& Ma 1983) and other previously developed methods in the literature, alongside a novel likelihood-based approach, which directly fits spectral and temporal decay features to IACT data. Through dedicated observation simulations, we evaluate the performance of these methods for the Large-Sized Telescope prototype (LST-1) of the Cherenkov Telescope Array Observatory (CTAO). We characterised the enhanced sensitivity found in some of these methods compared to the standard significance estimation. For GRBs simulated with very fast afterglow decay, an improvement of more than a factor of two is seen in detections for some of the methods presented here, compared to the standard Li\&Ma approach.}

\ConferenceLogo{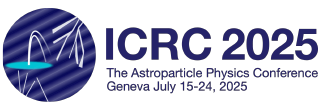}

\FullConference{39th International Cosmic Ray Conference (ICRC2025)\\
 15–24 July 2025\\
Geneva, Switzerland\\}
\begin{document}
\maketitle

%\section{ToDos}
%\begin{itemize}

%\item Plot of FAR as a function of method (leo gives edna the table and edna does the plot) 

%\end{itemize}
\section{Introduction}

Gamma-Ray Bursts (GRBs) are transient events from extragalactic sources. They can originate from the merging of two neutron stars or the core-collapse of a massive star. The signal is characterised by a prompt phase, lasting a few tens of milliseconds to hundreds of seconds, followed by an afterglow phase which can be detected in X-rays up to several weeks later. 

In the very-high-energy (VHE; 100 GeV–100 TeV) range, four GRBs have been detected by IACTs with the telescopes of the MAGIC and H.E.S.S. collaborations \cite{Abe_2023_1, Acciari_2019, Abdalla_2019, Abdalla_2021}, and one with the wide-field of view observatory, LHAASO \cite{lhaaso_2022}.

Imaging Atmospheric Cherenkov Telescopes (IACTs) operate in the VHE band but have a field of view of only a few degrees. IACTs rely on external alerts, mainly sent by X-ray and gamma-ray satellites, to point the telescopes in the direction of a GRB. Since the commissioning of the first Large-Sized Telescope (LST-1) in November 2019, one of the primary scientific objectives of the CTAO collaboration has been the detection of GRBs at VHE energies. To date, no GRB has been detected by the LST-1. 

Satellite detectors issue GRB alerts with typical latencies of tens of seconds, whereas the prompt emission can last from a few milliseconds up to tens of seconds. In most cases, LST observations must target the afterglow phase, which typically decays as a power law in time\cite{canonicalGRB}. The detection at VHEs is further complicated when considering the mean redshift of GRBs (z$\sim$2, \cite{Le_2017}), where absorption due to the extragalactic background light (EBL,\cite{domingues2011}) severely attenuates the emitted VHE photons. 

The LST-1 is an ideal instrument to mitigate these challenges~\cite{Acharya2017}. The LST-1 can reach any point in the sky in less than 30 seconds, minimising the observation delay of each GRB. Additionally, it has a low energy threshold of a few tens of GeV. At those energies, the EBL absorption is significantly reduced.  Even with these advantages, however, one still requires a robust statistical test to decide whether a genuine GRB signal has been recorded. The typical approach is the Li\&Ma method~\cite{LiMa}, which applies a likelihood ratio test, performed to assess the hypothesis of the absence of signal. To claim a detection, this hypothesis has to be rejected by more than $5 \sigma$.

The Li\&Ma likelihood‐ratio test is optimised for steady sources with constant flux but becomes suboptimal for GRB afterglows, whose signal decays as a power law. Averaging the signal over a typical two-hour LST-1 observation dilutes the early-time emission against a steady background, so a rapidly falling flux may fail to register as significant. 
In this contribution, we present alternative methods that can be more suitable to account for the temporal profile of GRBs, and we compare them to demonstrate that CTAO GRB detection ability can be improved.

\section{ Improved Statistical Methods for Detection of GRBs}
\label{sec:methods}

\subsection{Time-dependent Li\&Ma}

A first way to tackle the Li\&Ma limitation when considering decaying fluxes is to modify the likelihood formulation of Li\&Ma. This modification (Eq.~\ref{likelihood_tdep}), considers the time profile of the afterglow. The working principle involves dividing the likelihood into time bins, with either $0$ or $1$ event in each~\cite{tdeplima}. 

\begin{equation}
        \mathcal{L} = ( \prod_{t_i = (\Delta t,...,N \Delta t)} \frac{ (\Delta t (b + s(t_i) ) )^{ \{0,1\} } }{ \{0,1\}! } ) \cdot 
    \frac{ ( b \cdot T_{OFF} )^{N_{OFF}} }{ N_{OFF}! } e^{ -b T_{OFF} }
    \label{likelihood_tdep}
\end{equation}

The rest of the method is the same as Li\& Ma, constructing a ratio likelihood test to determine the level of significance. The signal is described as $s(t) =  \theta t^{-\alpha}$, with $\theta$ the amplitude, a free parameter, and $\alpha_a$, the (assumed) temporal index of the burst.
This method has already been used in VERITAS  to search for VHE emission in GRBs~\cite{Weiner:2015tva}.

\subsection{Temporal ON OFF}

The Temporal ON OFF method (tOn-Off) is described in \cite{hess_transient_method}. It adapts the Li\&Ma likelihood ratio test by comparing counts in a temporal ON region to those in the remainder of the temporal OFF region. 
To adapt this method to GRB afterglows, the ON region is set at the beginning of the simulated observation window, while the OFF region corresponds to the rest of the observation. Because the afterglow flux falls off rapidly, a sufficiently steep decay produces a count excess in the ON window compared to the OFF window, revealing the burst. Steady‐flux sources, by contrast, yield similar counts in both intervals and remain undetected under this method.

In this work, we try several ON region sizes using a geometric progression that doubles at each step, ranging from 0.469 to 30 minutes, with two longer-duration points added at 45 and 60 minutes. This is done to optimise the detection significance.

\subsection{ExpTest}

The ExpTest method~\cite{hess_transient_method, prahl1999} quantifies deviations of the inter-arrival time distribution from a Poisson process. An initial estimator is

\begin{align}
M &= \frac{1}{N} \sum_{\Delta T_i < C^\star} \Bigl(1 - \frac{\Delta T_i}{C^\star}\Bigr),
\end{align}

where \(N\) is the number of events, \(\Delta T_i\) the inter-arrival times, and \(C^\star\) their mean.  A normalised test statistic follows as

\begin{align}
M_r &= \sqrt{N}\,\frac{M - \bigl(e^{-1} - \frac{\alpha}{N}\bigr)}{\beta},
\end{align}

with \(\alpha,\beta\) from \cite{prahl1999}.  In practice, for each integration window at the start of the observation (spatial ON region), one computes \(M_r\) using the \(\Delta T_i\) within that window.  We further optimise detection by varying the length of this integration window over the predefined set of durations described in Sec.~3.2.

\subsection{CuSum test}

As in the case of ExpTest, this method relies on measuring the deviation of the interval time between events. Initially proposed in \cite{Page1954}, its adaptation for searching burst events is detailed in \cite{hess_transient_method}. The principle is to compute the cumulative sum of the time intervals $\Delta T_i$ between the arrival times $T_i$, in case of $N$ number of events (see Equation~\ref{CuSum}). 

\begin{equation}
    \chi_i = \sum_{k=1}^i \left( \Delta T_k - \frac{1}{N} \sum_{j=1}^N \Delta T_j \right)
    \label{CuSum}
\end{equation}

Then, by computing the variance (Equation~\ref{var_CuSum}), one can compute the number of standard deviations between the behaviour of the distribution $\Delta T_i$ and that of a steady source.

\begin{equation}
    Var\left( \chi_i \right) = \frac{i ( \frac{1}{N} \sum_{j=1}^N \Delta T_j )^2 }{N} \left( N - i \right)
    \label{var_CuSum}
\end{equation}

The time intervals $\Delta T_i$ are taken inside a test window time, defined at the beginning of the integration time. For this analysis, we used a slight modification of CuSum, in which the first and last ten events in the considered time interval are removed. This is done to eliminate edge-effect artefacts from the rapid initial decay and the low-rate tail of the afterglow, ensuring the CuSum test focuses on the central portion of the events where signal-to-background discrimination is most reliable.

\subsection{LifFT}

The Likelihood‐based Fitting for Transients (LiFT) method uses a likelihood‐ratio test between a null model (no source) and an alternative spectral–temporal model:

\begin{equation}
    \phi\left(E, t\right) = \phi_0 \left( \frac{E}{E_0} \right)^{-\Gamma} \left(\frac{t-t_{ref}}{t_0}\right)^{-\alpha},
    \label{lift}
\end{equation}

To improve stability, LiFT first fits the source model over all times (stacking the time axis) and then fits the temporal decay over all energies (stacking the energy axis); these intermediate results serve as seeds for the final joint fit.

%The scheme is detailed in Figure \ref{fig:lift}. 

%\begin{figure}[htbp]
%    \centering %remove figure and expand text explanation
%    \includegraphics[width=0.7\linewidth]{images/fitting_procedure.png}
%    \caption{Fitting procedure of the LiFT method}
%    \label{fig:lift}    
%\end{figure}

%The limitation of this method is its computing time. Indeed, because of the fitting procedure, it takes much more time than the other methods detailed before. Thus, it would not be usable in the Real Time Analysis (RTA).

\section{Testing procedure}

\subsection{Simulations}

To test all those methods, simulations have been performed, based on LST-1 instrument respose functions (IRFs)~\cite{garcia2022lstmcpipe, Abe_2023}%cite performance paper instead 
, using Gammapy-1.3 \cite{Gammapy_2025}. All the bursts were simulated with the same spectral and temporal description as in Eq.~\ref{lift}.  The photon index $\Gamma$ is set to $2.0$. We account for the absorption of VHE photons from the EBL by associating a redshift $z$ to each simulated GRB~\cite{domingues2011}, assuming four redshift values for the simulations (see column Redshift in Tab.~\ref{tab:table_param}), these values are chosen to test the low end of the redshift distribution of GRBs~\cite{Le_2017} in which EBL absorption is less dominant. We simulate several observation delays, which represent the elapsed time from the onset of the burst to the moment LST-1 begins its observations. To cover typical observation scenarios, we studied two cases (see Tab.~\ref{tab:table_param}): one with a short start-up delay of tens to hundreds of seconds, simulating a GRB that occurs during the night and is observed immediately, and one with a six-hour delay, representing a GRB that occurs during the day and is observed during the up-coming night (\emph{Delay} column in Tab.~\ref{tab:table_param}).
Based on the canonical GRB afterglow model~\cite{canonicalGRB}, we constrain the parameter space differently for each case: the early‐start scenario typically requires a steeper temporal decay. In contrast, the delayed‐start scenario is characterised by a more gradual decline several hours after the burst. The considered values of the temporal index are given in Table \ref{tab:table_param}. For the spectral model normalisation, we selected values by aligning our simulated spectra with LST-1’s sensitivity curves computed for short exposures. We chose endpoints that span from well above the sensitivity threshold—where detection is trivial—to well below it, where detection is effectively impossible. Intermediate normalisation levels were placed via logarithmic interpolation between these extremes. This range of normalisations ensures we can observe any sensitivity gains offered by our alternative methods. All these parameter values are summarised in Table \ref{tab:table_param}.

\begin{table}[hbtp]
\centering
\resizebox{\textwidth}{!}{%
\begin{tabular}{lllll}
\toprule
 & \textbf{Delay (s)} 
 & \textbf{Temporal index ($\alpha$)} 
 & \textbf{Normalization ($\mathrm{cm^{-2}\,s^{-1}\,TeV^{-1}}$)} 
 & \textbf{Redshift, z} \\
\midrule
\textbf{Early case} &
$[30.0,\ 95.0,\ 300.0]$ &
$[-1.0,\ -3.0,\ -5.0]$ &
\begin{tabular}[c]{@{}l@{}}$[2.00,\ 4.09,\ 8.37,\ 17.1,$ \\ $35.0,\ 71.7,\ 147,\ 300] \cdot 10^{-8}$\end{tabular} &
$[0.1,\ 0.4,\ 0.7,\ 1.0]$ \\
\midrule
\textbf{Late case} &
$[21600]$ &
$[-1.2]$ &
\begin{tabular}[c]{@{}l@{}}$[1.00,\ 1.76,\ 3.11,\ 5.48,$ \\ $9.65,\ 17.1,\ 30.0] \cdot 10^{-5}$\end{tabular} &
$[0.1,\ 0.4,\ 0.7,\ 1.0]$ \\
\bottomrule
\end{tabular}%
}
\caption{Parameter spaces used for simulating bursts in both early and late cases.}
\label{tab:table_param}
\end{table}

To test the methods, the aim is to simulate $\sim 1000$ bursts per combination of parameters. This leads to a total of $288,800$ simulations in the early case and $28,000$ in the late one. For each observation, we test the detection methods in several integration times. For the early case, we sample integration windows on a non-uniform grid: a fine, roughly doubling sequence at short times (0.5, 1, 2.5, 5, 10 min), followed by coarser 30 min steps out to 240 min (30, 60, 90, 120, 150, 240 min). For the late case, we use the same pattern, with short times (10, 15, 22.5, 30, 37.5, 45, 60 min) followed by 30 min steps (90, 120, 150, 180, 210, 240 min). 

\subsection{Optimisations}

As described in Sec.~\ref{sec:methods}, to optimise each method, the ExpTest, CuSum and tOn-Off tests are performed using several time windows. In a similar way, for the case of time-dependent Li\&Ma we tested different temporal decay index assumptions. For each of these cases, to select the best set of optimisation parameter values, we calculate the true-positive rates (TPR) and false-alarm rates (FAR) from our simulated bursts, and turn those into a figure of merit (FoM = TPR\,-\,FAR). We then pick the optimisation value that gives the highest FoM. Performing this analysis at every integration time yields the optimal parameter curve, which was used to generate our detection fraction results presented in Sec.~\ref{sec:results}.

\section{Results}
\label{sec:results}

\begin{figure}[htbp]
    \centering 
    \includegraphics[width=0.99\linewidth]{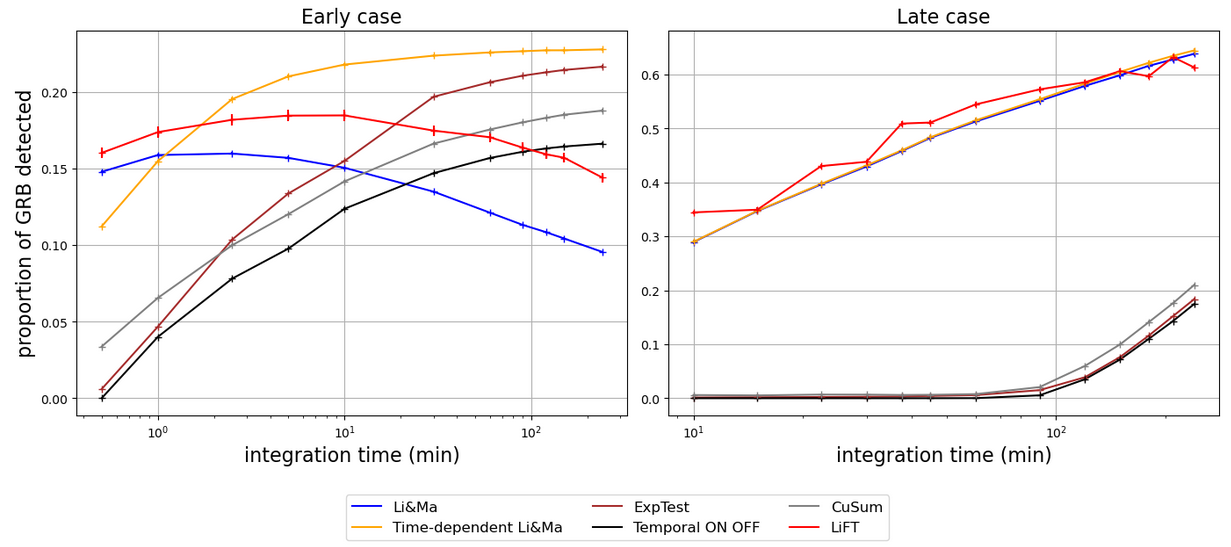}
    \caption{Proportion of GRBs detected with the different methods depending on the integration time, with errorbars, in early and late cases}
    \label{fig:results}    
\end{figure}

To compare the performances of the different methods applied to both cases, one can study the proportion of GRBs detected, depending on the integration time (see Fig.\ref{fig:results}). As explained above, ExpTest and tOn-Off have been optimised for the ideal window time of each method. In the curves shown in Fig.~\ref{fig:results}, this optimised time window is used to compute the proportion of detected GRBs at each integration time. For ExpTest the ideal time window corresponds to 0.47 min for integration times below 10 min and 0.93 min for integration times above 10 min. For tOn-Off the optimal window time is of 0.47 min for integration times below 5.0 min and 1.875 min for integration times above 5.0 min. 

%\begin{figure}[htbp]
%    \centering
%    \begin{subfigure}[t]{0.49\textwidth}
%        \centering
%        \includegraphics[width=\linewidth]{images/propDet_late.png}
%        \caption{Late observation}
%        \label{fig:result_late}
%    \end{subfigure}
%    \hfill
%    \begin{subfigure}[t]{0.49\textwidth}
%        \centering
%        \includegraphics[width=\linewidth]{images/propDet_early2.png}
%        \caption{Early observation}
%        \label{fig:result_early}
%    \end{subfigure}
%    \caption{Proportion of GRBs detected with the different methods depending on the integration time}
%   \label{fig:result_both}
%\end{figure}

\begin{figure}[h]
    \centering
    \includegraphics[width= 0.99\linewidth]{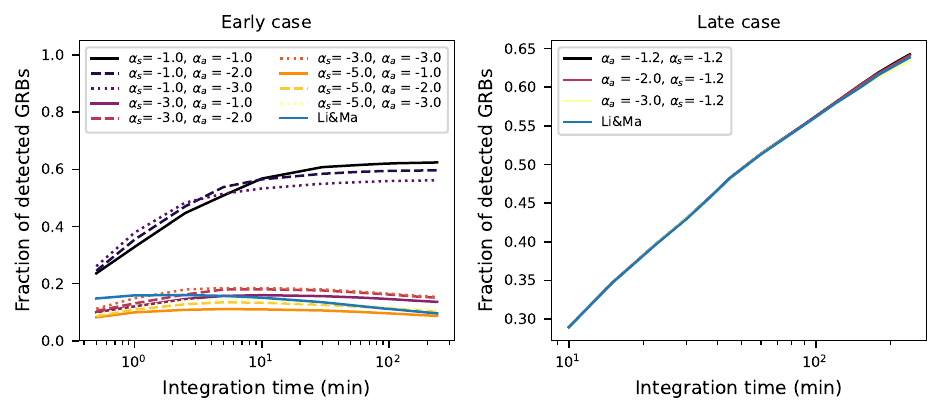}
    \caption{Time-dependent Li\&Ma: Fraction of simulated GRBs detected above a 5$\sigma$ threshold as a function of integration time, for both the early (left) and late (right) observing cases. In each panel, the coloured curves correspond to different simulated temporal indices $\alpha_s$, while the line styles indicate the assumed analysis temporal index $\alpha_a$; the solid blue curve shows the Li\& Ma detection method for comparison. }
    \label{fig:tdeplima}
\end{figure}

\subsection{Early case}

The right panel of Fig.~\ref{fig:results} illustrates the early observation scenario, in which the afterglow decay is particularly steep. The standard Li\& Ma detection fraction grows with integration time up to a point and then declines as the signal is buried in background. The time‐dependent Li\& Ma algorithm achieves the highest detection efficiency, matching the performance of the temporal ON–OFF and CuSum methods, and delivers twice the detection fraction of the standard Li\& Ma approach. 

In Fig.~\ref{fig:tdeplima}, we contrast the performance of tdep-Li\&Ma for different combinations of assumed and simulated temporal decay indices ($\alpha_a$ and $\alpha_s$, respectively). This suggests that one can enhance the tdep-Li\&Ma detection efficiency if $\alpha_a$ is closer to or equal to $\alpha_s$. 

For this simulation case, we computed the FAR of each method averaged over all integration times and found all of them have FAR values lower than 0.1\%, including Li\&Ma and tdep-Li\&Ma with FAR=0.

\subsection{Late case}

The left panel of Figure \ref{fig:results} shows the late‐case results. As anticipated, the Li\&Ma method and its time‐dependent variant achieve the highest detection fractions. Because the signal decays more slowly in this scenario, techniques optimised for rapid variability are comparatively less efficient, requiring much longer integration times before they register even a small fraction of the simulated bursts. This is because the slow, extended emission allows integrated-count methods, such as tdep-Li\& Ma, to steadily build up the signal-to-noise ratio. In contrast, rapid-variation algorithms fragment the flux into low-SNR segments. This “late” case thus serves as a useful sanity check, confirming that each method behaves in line with theoretical expectations.

In contrast with the early case, Fig~\ref{fig:tdeplima} indicates that tdep-Li\&Ma practically becomes insensitive to the choice of $\alpha_a$. This behaviour is expected, as for such late times, the flux decays so slowly that the assumed decay does not affect the relative time weights of the signal versus background (see Eq.~\ref{likelihood_tdep}). 

\section{Conclusion/Discussion}

Overall, our study demonstrates that incorporating explicit temporal models significantly enhances the GRB detection capabilities of LST-1, achieving high sensitivity while maintaining low FARs (all methods except CuSum satisfy a $<$0.1\% false‐alarm requirement).
Given their relatively simple structure, Li\&Ma, tdep-Li\&Ma, CuSum, ExpTest, and tOn-Off are methods that do not require long computing times; therefore, they are suitable for obtaining initial results on observed GRBs during the night with real-time analysis (RTA), which is planned. As an exception to this is  LiFit, which performs a full likelihood fit and can take several hours of computing runtime; however, it provides better performance at short integration times in the early case. 

From our benchmarks, the tdep-Li\&Ma test yields the highest detection fractions by optimally weighting counts according to an assumed power-law decay index; in the RTA context, one can adopt canonical values (e.g.\ –3.0 for early-start cases, –1.2 for late-start cases) and then refine them offline once the X-ray light curve is available.

A key limitation of this work is the assumption of a constant background rate, using IRFs fixed at a 20\degree ~zenith angle over a two-hour integration. For bursts near transit, this approximation introduces only minor errors. Accounting for background variations will be important for off-zenith or longer observations.

Finally, integrating one or more of these rapid, time‐aware algorithms (alongside the standard Li\&Ma test) into the LST-1 real-time pipeline would provide continuously updated significance estimates and support informed decisions on extending or terminating GRB follow-up during observing nights.

\textbf{Acknowledgements} This work has been done thanks to the facilities offered by the Univ. Savoie Mont Blanc - CNRS/IN2P3 MUST computing center. We thank the LST-CTAO Collaboration for providing the Instrument Response Functions used in this study and for fruitful discussions.

\end{document}